\RequirePackage{amsmath}
\documentclass[11pt]{iopart}
\usepackage{amsfonts,amssymb,color,amstext}
\newcommand{\ket}[1]{{\left| {#1} \right\rangle}}
\newcommand{\bra}[1]{{\left\langle {#1} \right|}}

\newcommand{\C}{\mathcal{C}}

\usepackage[pdfpagelabels,pdftex,bookmarks,breaklinks]{hyperref}
\definecolor{deeppurple}{RGB}{100,0,120} 
\definecolor{darkgreen}{RGB}{0,150,0}
\definecolor{darkblue}{RGB}{0,0,130}
\hypersetup{colorlinks=true, linkcolor=deeppurple, citecolor=darkgreen, filecolor=red, urlcolor=darkblue}
\hypersetup{pdftitle={Programmable Quantum Simulation by Dynamic Hamiltonian Engineering}, pdfauthor={David Hayes, Steven T. Flammia, Michael J. Biercuk}}

\usepackage{iopams}
\usepackage{graphicx}
\begin{document}

\title{Programmable Quantum Simulation by Dynamic Hamiltonian Engineering}

\author{David Hayes}
\address{ARC Centre for Engineered Quantum Systems, School of Physics,\\ The University of Sydney, NSW 2006 Australia}
\address{National Measurement Institute, Bradfield Road, West Lindfield, NSW 2070 Australia}

\author{Steven T.\ Flammia}
\address{ARC Centre for Engineered Quantum Systems, School of Physics,\\ The University of Sydney, NSW 2006 Australia}

\author{Michael J.\ Biercuk}
\address{ARC Centre for Engineered Quantum Systems, School of Physics,\\ The University of Sydney, NSW 2006 Australia}
\address{National Measurement Institute, Bradfield Road, West Lindfield, NSW 2070 Australia}
\ead{michael.biercuk@sydney.edu.au}

\begin{abstract}
Quantum simulation is a promising near term application for quantum information processors with the potential to solve computationally intractable problems using just a few dozen interacting qubits~\cite{PhysRevA.65.042323, PhysRevLett.82.5381, Blatt:2012}. A range of experimental platforms have recently demonstrated the basic functionality of quantum simulation applied to quantum magnetism, quantum phase transitions, and relativistic quantum mechanics ~\cite{PhysRevA.61.012302, Kim:2010, Edwards:2010,Gerritsma:2011, Barreiro:2011, Lanyon:2011, Bloch:2012, Schindler:2013}. However, in all cases, the physics of the underlying hardware restricts the achievable inter-particle interactions and forms a serious constraint on the versatility of the simulators.  To broaden the scope of these analog devices, we develop a suite of pulse sequences that permit a user to efficiently realize average Hamiltonians that are beyond the native interactions of the system~\cite{Leung:1999, Leung2002}.  Specifically, this approach permits the generation of all symmetrically coupled translation-invariant two-body Hamiltonians with homogeneous on-site terms, a class which includes all spin-$1/2$ XYZ chains, but generalized to include \emph{long-range} couplings. Our work builds on previous work proving that universal simulation is possible using both entangling gates and single-qubit unitaries~\cite{PhysRevA.65.042309, PhysRevA.65.040301, QIC2002}.  We show that determining the appropriate ``program'' of unitary pulse sequences which implements an arbitrary Hamiltonian transformation can be formulated as a linear program over functions defined by those pulse sequences, running in polynomial time and scaling efficiently in hardware resources. Our analysis extends from circuit model quantum information to adiabatic quantum evolutions, representing an important and broad-based success in applying functional analysis to the field of quantum information.
\end{abstract}

\maketitle

\section{Introduction}
Universal computation is derived from the ability to reduce complex and arbitrary calculations to a finite set of physical operations. While generic digital logic is well suited for many applications, \emph{analog} simulation can provide much more efficient methods for performing certain types of computations. In this approach, computational operations of interest can be efficiently built into the underlying hardware, for instance, as analog circuits.  Such approaches can greatly enhance performance at the cost of guaranteed error tolerance~\cite{MacLennan:2007}. 

The same capabilities exist in the quantum regime, with the promise to realize extraordinary computational capabilities for a variety of challenging problems of interest.  Analog quantum simulation may provide a near-term path to providing solutions to currently intractable computations at the scale of just a few dozen interacting quantum systems.  Quantum simulation leverages well-controlled quantum coherent devices in order to study the properties of poorly understood many-body systems: e.g. interacting spins and quantum magnetism.  In this vein, a wide variety of experimental demonstrations have been realized to date~\cite{Kim:2010, Edwards:2010, Blatt:2012, Bloch:2012, Lanyon:2011} using precisely controlled atomic systems.

The efficiency benefits associated with analog simulators are partially offset by the challenging open question as to how one may efficiently implement a ``program'' therein using a constrained set of basic resources.  This is apparent in the case of experimental simulation of quantum magnetism; all experiments to date have been severely constrained by the native interactions achievable between the underlying quantum bits. Attempts to expand functionality have typically focused on exploiting the specific properties of a given hardware platform~\cite{AspuruGuzik:2012, Korenblit:2012} (e.g. varying the coupling of spins to different normal modes of motion in a trapped-ion simulator), rather than providing a generic, technology-independent approach.  Producing a general, efficient, and extensible framework for programming complex interactions and broadening the range of accessible simulations with limited hardware capabilities is thus a key requirement to expand the utility of mesoscale quantum simulation.

Here we describe a technology-independent technique to develop a hybrid quantum spin simulator with constrained resources, capable of efficiently realizing a broad class of interacting spin Hamiltonians.   Beginning with a native long-range pairwise interaction between spins on a one-dimensional chain, the application of short sequences of single-spin Pauli operators can effectively generate any other Hamiltonian in the family by filtering the time-averaged relative weight of different pairwise interactions.  The approach we describe frames the problem as one of functional synthesis where basis functions are generated by the pulse sequences that we define. This general methodology has met success in a variety of contexts for quantum information, including the design of algorithms for quantum memory~\cite{HayesPRA2011, LongStorage}, robust-control techniques~\cite{OwrutskyArXv2012, Soare2014}, circuit-model algorithm synthesis~\cite{Welch2014}, and even applications in quantum-enabled sensing~\cite{Cooper14}.

Through sequential and concatenated application of the pulse sequences we define, the basis functions can be added and multiplied together and are shown to generate a complete, orthogonal set of functions that span the full space of Hamiltonians in the family. The challenge of determining the appropriate operations required to map the native interaction to any other is reduced to an efficiently soluble linear program, and moreover this solution is (in a certain precise sense) optimal. In addition to providing specific examples of useful filtering tasks, we show how this technique is extensible to adiabatic quantum simulation and permits the preparation of states that are otherwise inaccessible.   

Specifically, we consider the family of one-dimensional Hamiltonians with translation-invariant symmetrically coupled two-body terms (i.e. $XX$-, $YY$-, and $ZZ$-type couplings) extended to include long-range couplings, and homogeneous transverse on-site terms.  More precisely, we model a finite $N$-qubit chain on a one-dimensional lattice (Fig.~\ref{Fig:F2}a) with Hamiltonian $H = H_I + H_T$. Here $H_{I}$ is a fully connected interaction Hamiltonian
\begin{equation}\label{eq:Hamiltonian_aniso}
	H_I=\sum_{d=1}^{N-1}\Omega_d \hat{h}_{d}=\sum_{d=1}^{N-1}\mathbf{\Omega}_{d} \cdot \sum_{j=1}^{N-d-1}\mathbf{S}_j\mathbf{S}_{j+d}\,,
\end{equation}
where $\mathbf{\Omega}_d=(\Omega_{d}^{(x)},\Omega_{d}^{(y)},\Omega_{d}^{(z)})$ and $\mathbf{S}_j\mathbf{S}_{j+d}=(X_j X_{j+d},Y_j Y_{j+d},Z_j Z_{j+d})$ represent the interaction strength and coupling in each Cartesian direction of two qubits separated by $d$ lattice spacings. The dot product between these implements a generally anisotropic coupling. We also incorporate a homogeneous transverse field $H_T = B \sum_j Z_j $, where the magnitude of $B$ is controllable. 

Our first main result is that, starting with any such family with polynomially decaying long-range couplings $\Omega_d$, there exists a short sequence of single-qubit $Z$ pulses such that the time-averaged Hamiltonian is equal to \emph{any other Hamiltonian in the class}, up to rescaling factor. We can quantify this statement as follows. The sequence of $Z$ pulses contains at most $O(\log N)$ concatenation steps, requiring a pulse number of at most $O(N^2)$ with at most $O(N^3)$ individual spin flips, and the rescaling factor for the filtered interaction strength is at most $N^{-2.585}$, the exponent given by $-\log_2 6$. Thus the scheme can be fully implemented in polynomial time. Moreover, the \emph{optimal} rescaling factor for our scheme can be obtained by efficiently solving a linear program. We also provide explicit demonstrations of our method for implementing power-law decay modulation in Ising spin chains and adiabatic state preparation, as well as explicit bounds on simulation error.

The remainder of this paper is organized as follows. We describe our general approach to Hamiltonian engineering in Sec.~\ref{S:Heng}, introducing an extensible family of pulse sequences that allow us to generate any Ising-type coupling (in the above sense).  Through combination of filters we demonstrate how these pulse sequences are in fact universal, allowing arbitrary mappings within the Hamiltonian class with efficient resource scaling in Sec.~\ref{S:CN}, with additional proof details in~\ref{A:CN}. We then show how filters may be efficiently compiled to achieve a desired ``program'' for a quantum simulator in Sec.~\ref{S:lp}, followed by examples of what can be achieved with the method in Sec.~\ref{S:examples}. We then show how this can all be generalized to adiabatic evolutions (Sec.~\ref{S:Adiabatic}) and consider the fully general Hamiltonians of Eq.~(\ref{eq:Hamiltonian_aniso}) in Sec.~\ref{S:Heisenberg}, including rigorous bounds on the simulation error. Our work concludes with a discussion of open questions in the literature and a summary of our results.

\section{Hamiltonian Engineering}\label{S:Heng}

\subsection{General Concepts}
Owing its foundations to research in NMR \cite{Levitt:1986, PhysRevA.61.012302}, the idea of altering the dynamics of interacting quantum systems has proven invaluable in the field of quantum control. For instance, in the context of dynamic error suppression~\cite{Viola:1998, Biercuk:2009, Ng:2011, Peng:2011} a quantum system may be effectively isolated from its environment, not by eliminating the physical interaction, but by inducing a dynamical response which acts to average out the coupling to the environment. This picture extends to interactions between quantum coherent systems as in spin decoupling and recoupling in NMR~\cite{PhysRevA.61.012302, Leung:1999, Leung2002, PhysRevA.64.052301} and have been used to provide universality proofs of quantum simulators using a set of operations similar to universal quantum computation~\cite{PhysRevA.65.042309, PhysRevA.65.040301, QIC2002}.

The simplest example of this type of Hamiltonian engineering is the dynamic decoupling of two spins interacting via the Hamiltonian $H=\Omega X_1X_2$, where $X_j$ is the Pauli $x$ operator for the $j$th particle.  The application of phase-flip pulses ($Z_{j}$) at time intervals $\Delta t$ produces time-dependent operators $X_{j}(t)=w_{j}(t) X_{j}$ in the Heisenberg picture.  The control propagator $w_{j}(t)=\pm 1$ captures the discrete-time modulation with the sign changes occurring at the times of the applied pulses and time evolution operator is,
\begin{equation}
	U(\Delta t)=\exp\bigl[-i\Omega_{\mathrm{eff}}X_1X_2 \Delta t \bigr] \,,
\end{equation}
\noindent with $\Omega_{\mathrm{eff}}=\Omega \textbf{w}_1\cdot\textbf{w}_2$ and $[\textbf{w}_j]_l=w_j(l \Delta t)$.  We thus see that the inner product of the discretized control propagators $\textbf{w}_j$ determines the effective coupling.

It has been demonstrated that the physics underlying this simple example can be generalized for application to a collection of qubits in order to realize arbitrary effective two-qubit interactions~\cite{Leung:1999}. Unfortunately existing approaches entail tremendous complexity when employed for Hamiltonian engineering in quantum simulations.  Our goal is to produce a simple extensible set of pulse-sequence constructions suitable for the task of quantum simulation. Our method takes advantage of the symmetry of a ubiquitous architecture in physical quantum simulators, the translational invariant one-dimensional spin chain with long-range coupling.

\begin{figure*}[tb]
\centering
\includegraphics[width=14cm]{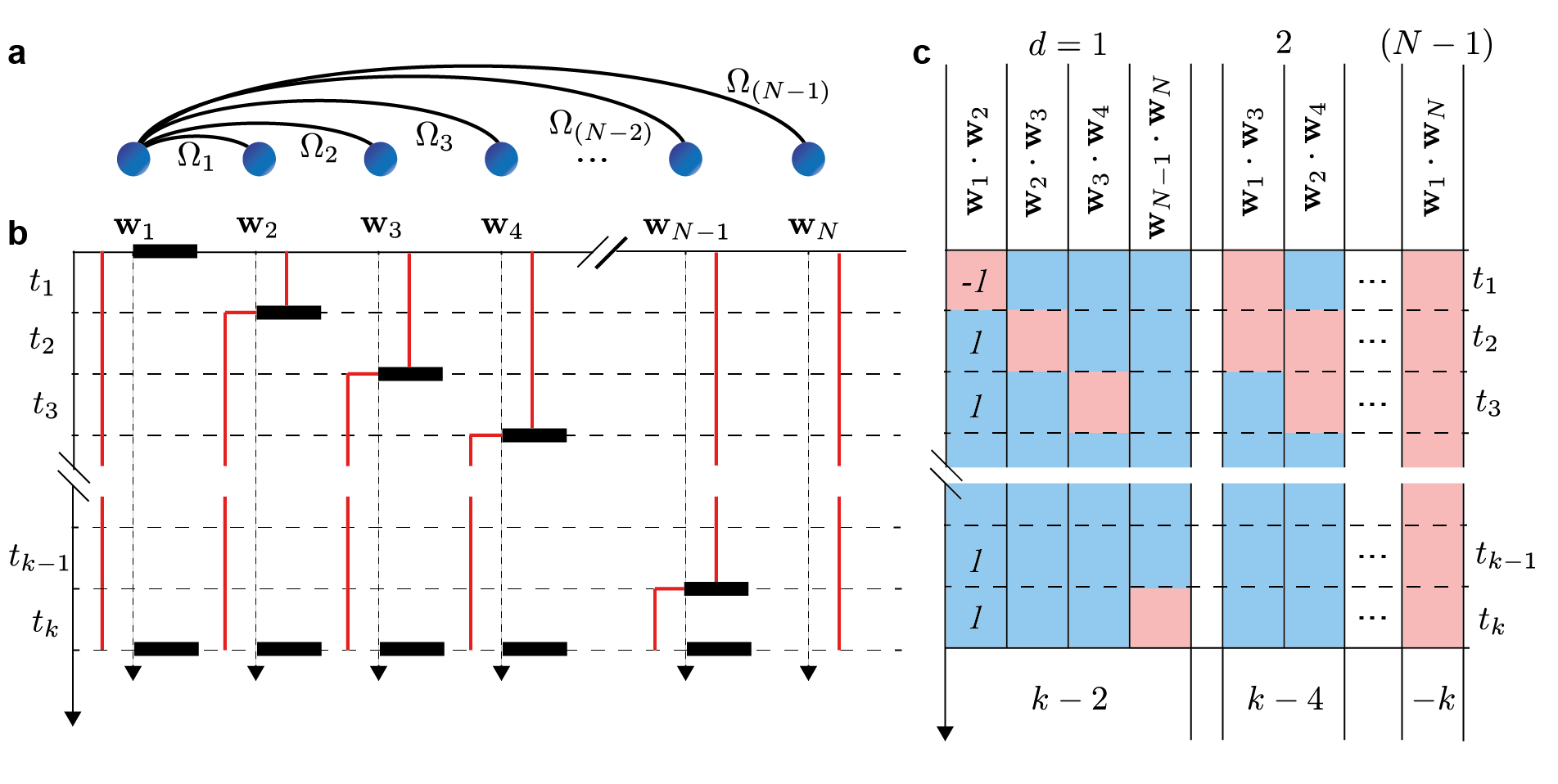}\\
\caption{(color online) Illustration of translationally invariant dynamical filter generation for a linear chain of coupled qubits.  a)~Schematic depiction of qubits on a linear chain with distance-dependent coupling.  For an $N$-qubit chain the largest distance is $(N-1)d$, where $d$ is the primitive lattice constant.  Only interactions between the qubit 1 and others in the chair are presented for clarity.  b)~Pulse sequence and associated control propagators, $\{\mathbf{w}^{(k)}_{j}\}$, for the $N$ qubits. Time runs from top to bottom in $k$ discrete steps.  Black bars indicate $\pi_{Z}$ pulses, causing a sign change in the associated control propagator.  The particular map depicted generates $f^{\Lambda}_{k=N}(d)$.   c)~Graphical representation of the vector dot-products, $\mathbf{w}^{(k)}_{j}\cdot\mathbf{w}^{(k)}_{j'}$ between spins entering into the time evolution operator.  Color indicates the sign of the interaction in a discrete time bin.  The sum over time bins gives the resulting prefactor in the effective Hamiltonian.  All pairs of spins separated by distance $d$ have a resulting overlap $k-1$, all pairs separated by $2d$ have overlap $k-2$, decreasing linearly with $d$ until the qubits separated by the largest distance have overlap $-k$, indicating a sign change.  
\label{Fig:F2}}
\end{figure*}

\subsection{Filter construction}
We now show how to build on this very simple example and define a suite of pulse sequences that may be used to efficiently program an effective long-range pairwise spin-coupling Hamiltonian in a multi-qubit system. Consider a Hamiltonian of the form in Eq.~(\ref{eq:Hamiltonian_aniso}); this form of long-range spin interaction is extremely important for quantum-many-body physics studies of exotic forms of quantum magnetism and is commonly realized e.g.\ in trapped-ion spin simulators and optical lattice simulators~\cite{Lloyd:1993,Brennen:2000,Saffman:2010, Kim:2010, Britton:2012}.

We begin by treating the special case of Ising-type interactions,
\begin{equation}
	H_I\to\sum_{d=1}^{N-1}\Omega_d \hat{h}_{d}=\sum_{d=1}^{N-1}\Omega_{d}\sum_{j=1}^{N-d-1}X_jX_{j+d}\,,
\label{eq:Hamiltonian}
\end{equation}
\noindent and determine the relevant pulse sequences that permit our desired transformations on the interqubit couplings to be realised exactly.  Later we will revisit the generalised anisotropic Heisenberg-type coupling with homogeneous on-site terms and discuss what modifications must be made in order to effectively treat the more general model.  We look to craft a set of pulse sequences that permit the realization of an arbitrary effective Hamiltonian within this class by modifying the native interaction so that $\Omega_d\rightarrow\Omega_{d}f(d)$.  In this approach both the original and effective Hamiltonians are translationally invariant. 

To simulate the time-evolution under a Hamiltonian for a time $T$, we break $T$ into $k$ segments of equal duration $\Delta t_k = T/k$, with pulses applied as required at the transitions between time segments~\cite{PhysRevA.65.040301}. This results in a family of control propagators $\{\textbf{w}^{(k)}_{j}\}$ labelled by $k$ with $j$ indexing the qubits. We define operators $\hat{P}_{k,l}$ that are applied both at the beginning and end of the $l$th time segment\footnote{Thus at the boundary of time segments $l$ and $l+1$ we actually apply the operator $\hat{P}_{k,l} \hat{P}_{k,l+1}$. Defining $\hat{P}_{k,l}$ in this way greatly simplifies the analysis.}. Each is a product of $Z_{j}$ operators on some subset of the $N$ different qubits in the chain. The result of applying these pulse sequences is a modification of the time evolution during the $l$th time segment to $U_{k,l}(\Delta t_k)=\hat{P}_{k,l}\exp\bigl[-i H \Delta t_k\bigr]\hat{P}_{k,l}$. The $\hat{P}_{k,l}$ operators generate sign changes in the control propagators of individual qubits and the sign of $\hat{P}_{k,l}X_j\hat{P}_{k,l}$ is encoded by the $l$th element of $\textbf{w}^{(k)}_{j}\equiv [\textbf{w}^{(k)}_{j}]_{l}$ (Fig.~\ref{Fig:F2})b-c. The effective time-evolution is related to the individual control propagators as $U_{k,l}(\Delta t_k)=\exp\bigl[-i H^{\mathrm{eff}}_{k,l} \Delta t_k\bigr]$ where $H^{\mathrm{eff}}_{k,l}=\sum_d \Omega_{d} \sum_j[\textbf{w}^{(k)}_{j}]_l[\textbf{w}^{(k)}_{j+d}]_{l}X_{j}X_{j+d}$. The total evolution operator is given by the product over all time periods as
\begin{align}
	U_k(T)=\prod^{k}_{l=1}U_{k,l}(\Delta t_k)=\exp\bigl[-i H_k^{\mathrm{eff}} T \bigr],\label{eq:ProductOfUnitaries}\\
	H_k^{\mathrm{eff}}=\sum_lH^{\mathrm{eff}}_{k,l} \equiv \sum_d \Omega_{d}f_k(d) \hat{h}_d
\end{align}
%
The dynamically modified coupling is represented through the function $f_k(d)$, which we define to be the vector dot product of the control propagators $\textbf{w}^{(k)}_{j}\cdot\textbf{w}^{(k)}_{j+d}$ over the discrete time segments. 

\begin{figure}[htp]
\centering
\includegraphics[width=8cm]{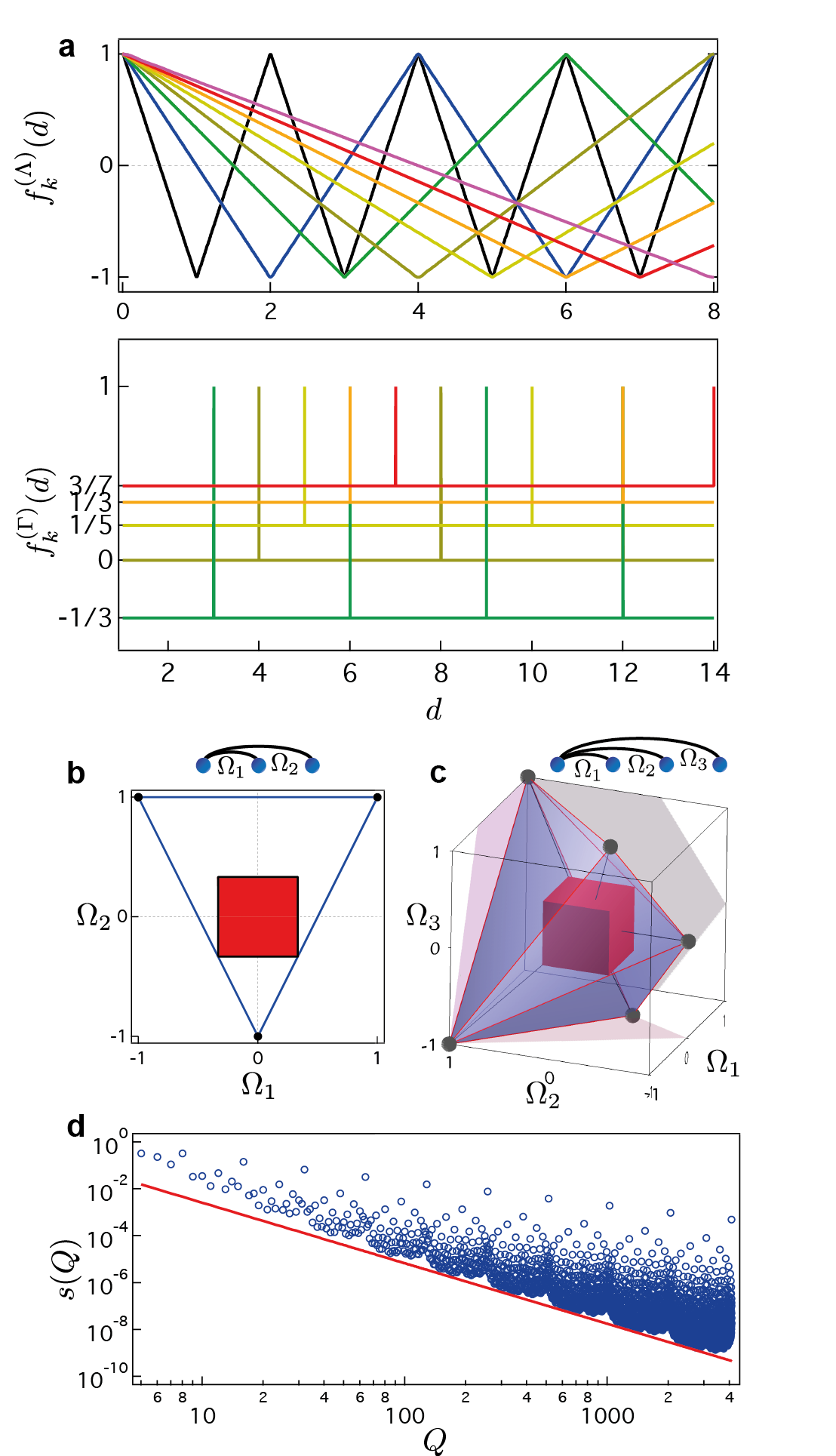}\\
\caption{(color online) Filters achievable using dynamical mapping pulse sequences. a) Upper Panel: Basis of triangle wave functions arising from $f^{\Lambda}_{k}(d)$.  Traces plotted here have $k=1$ (black, highest frequency) to $k=8$ (violet, lowest frequency).  Lower Panel: Basis of boost filters arising from $f^{\Gamma}_{k}(d)$, for $k=3$ to $k=6$ using the same color encoding as in the upper panel. b) A representation of the set $\C_3$ and the universal filter space contained within it, represented by the square confined in size to fit within the convex set defined along axes $\Omega_{d=1}$ and $\Omega_{d=2}$, the only possible values for three spins.  Black points represent the extremal values of accessible filters, $(1,1)=\Lambda_0$, $(-1,1)=\Lambda_1$ and $(0,-1)=\Lambda_2$ in the two dimensions, bounding a convex set denoted by the blue line.  Only $\Lambda_0$ and $\Lambda_2$ are required to construct any other triangle wave where $k\geq 2$, since all higher order waves are a weighted average of these two maps when $d\leq 2$, and concatenation of sequences does not yield new filters:  $\Lambda_2^2=(\Lambda_0+\Lambda_1)/2$, $\Lambda_1^2=\Lambda_0$ and $\Lambda_1\Lambda_2=\Lambda_2$.  c) A representation of the set $\C_4$ and the universal filter space (cube) contained within it.  Projections of the complex polytope onto planes in the basis of pairwise qubit coupling are represented using shading.  In b) and c) schematic representations of possible interqubit couplings (giving axes on the filter space) are presented.  d) Calculated value of $s(Q)$, the strength of the filtered interaction for qubits separated by distance $Q>N/2$.  Red line is a guide to the eye scaling as $Q^{-2.585}$.
\label{Fig:F3}}
\end{figure}

With this framework we introduce the primary class of pulse sequence we construct, $\Lambda_k$, defined by the pulse operators
\begin{equation}
	\Bigl[\hat{P}_{k,j}\Bigr]\Bigr |_{j=1}^{k}\;\;\;\;\;\hat P_{k,j} = \prod_{i=1}^NZ_i^{\bigl\lfloor \frac{i-j-1}{k}\bigr\rfloor}.
\end{equation}
Here, square brackets around the operators $P_{k,j}$ indicate that we are defining a set of operators acting simultaneously on all qubits in the chain. We also denote rounding $x$ down (respectively, up) to the nearest integer by $\lfloor x \rfloor$ (respectively, $\lceil x\rceil$).  In this construction the exponent on the $i$th Pauli $Z$ operator is always an integer so that, depending on the parity, either $Z_i$ or the identity is applied to the $i$th qubit and $j$ denotes the relevant time bin.  These sequences modify $\Omega_d$ by the factor 
\begin{equation}
	f^{\Lambda}_{k}(d):= (-1)^{\left\lfloor \frac{d}{k}\right\rfloor}\left(1-2\left\{\frac{d}{k}\right\}\right)\,,
\end{equation}
where $\left\{x\right\}$ denotes the fractional part of $x$. This filter varies as a triangle wave as a function of qubit distance (as shown in Fig.~\ref{Fig:F3}a), with period $2k$ and range $f^{\Lambda}_{k}(d)\in[-1,1]$ which allows for considerable flexibility in the effective coupling. For example, it is possible to switch the sign of the interaction from ferromagnetic to antiferromagnetic, or visa versa,  as a function of distance, $d$.

The second class of filters we introduce is defined as $\Gamma_k$:
\begin{equation}
	\Bigl[\hat{P}'_{k,j}\Bigr]\Bigr |_{j=1}^{k}\;\;\;\;\;\hat {P'}_{k,j} = \prod_{i=1}^{N}Z_{i}^{\left\lceil \left\{ \frac{k+i-j}{k} \right\} \right\rceil }.
\end{equation}	
The filter $\Gamma_k$ is capable of providing a relative enhancement to the interaction strength of qubits separated by integer multiples of $k$ lattice spacings. The $\Gamma_k$ map modifies the system's evolution according to 
\begin{equation}
	f^{\Gamma}_{k}(d)=1-\frac{4}{k}\biggl\lceil \Bigl\{\frac{d}{k} \Bigr\}\biggr\rceil \,.
\end{equation}
\noindent Because of the selectivity of the filter's action in providing a periodic (in $d$) relative enhancement in coupling, we refer to this as a boost filter.

\section{A complete set of interactions}\label{S:CN} 
The simulation of arbitrary Hamiltonians mandates much more complex functional dependences than simple triangle waves or periodic boosts.  Previous work~\cite{Green1979} has shown that triangle waves coupled with phase shifts form a complete basis for functional synthesis similar to sines and cosines.  Despite being unable to generate phase shifted triangle waves, we show that it is possible to achieve arbitrary multiqubit interactions when we augment positive linear combination of filters with products of filters.  Utilizing both operations overcomes the limitations arising from accessing only positive linear combinations and ``cosine''-like triangle waves.  

\subsection{Universal filter space}
A positive linear combination such as, $\Omega_d\left[T^{(1)}f^{(1)}(d)+T^{(2)}f^{(2)}(d)\right]$, is generated through sequential application of the pulse sequences with the relative weights being determined by the time taken to implement the pulse sequence; cf.\ Eq.~\ref{eq:ProductOfUnitaries}. Similarly, products of filters such as $\Omega_d\left[T^{(1)}T^{(2)}f^{(1)}(d)f^{(2)}(d)\right]$ are achieved via sequence concatenation, in which sequences are ``nested'' within the free evolution periods of one another.  These two operations are simple but extremely powerful;  with these it is possible to engineer \emph{arbitrary} translationally invariant qubit couplings within the family we consider.

Achieving this among $N$ qubits implies the ability to perform functional synthesis in an $(N-1)$-dimensional space, where each dimension represents the coupling between qubits separated by a particular distance, $d$. A point in this space therefore represents an arbitrary combination of interaction strengths over the relevant values of $d$.  In this space, we define $\C_N$ to be the convex set of filters $f_{k}(d)$ generated by both positive linear combination and concatenation of our basic filters. While this set takes a complex geometric form, we will show that it contains an $(N-1)$-dimensional hypercube around the origin, implying that these filters achieve an arbitrary $f(d)$ and hence an arbitrary $\Omega_{\text{eff}}(d)$.   We call this cube ``universal'', in analogy with universal computation, but here restricted to translation-invariant two-body spin Hamiltonians with on-site transverse fields. In Figure~\ref{Fig:F3}, we show $\C_3$ and $\C_4$, including the universal set contained therein.  
\subsection{Efficient Resource Scaling of the Method}
Proving access to a universal filter space still leaves open the question of how the actual control operations affect the scaling of the simulator's performance.  We prove that it is possible to efficiently generate an arbitrary desired $\Omega_{\text{eff}}(d)$ using linear combination and concatenation of the primary sets of filters we define.  A key result of our method, described in detail in~\ref{A:CN}, is that the realization of an arbitrary interaction Hamiltonian within this universal space may be accomplished using at most $O(\log N)$ concatenation steps. These are the most resource-intensive operations we employ, requiring a pulse number of at most $O(N^2)$ with at most $O(N^3)$ individual spin flips.   Arriving at this result is a complex exercise, and finding such an efficient resource scaling was not expected at the outset of this work.

Our ability to efficiently construct the universal filter space has significant impact on the \emph{runtime} of filter implementation.  The cost of universality is a rescaling of the interaction strength, reflected in the reduced volume of the hypercube relative to the extremal points of the convex set (see Fig.~\ref{Fig:F3}b-c).  This reduced effective interaction strength within the universal filter space physically corresponds to an increase in the total evolution time required to achieve a target interaction form.

Naively, requiring $O(\log N)$ layers of concatenation in realizing the universal filter space would result in a bound of $N^{O(\log N)}$, which is only quasi-polynomial.   To test this we calculate the quantity $s(Q)$, the ``worst-case'' strength of the filtered interaction for qubits separated a distance $Q>N/2$, whose inverse sets the upper bound on runtime
\begin{equation}
	s(Q) = \prod_{n=0}^{m}\left(f^{\Lambda}_{2^{n}}+\frac{1}{(1-\frac{4}{Q})}\right)\,.
\end{equation}
(see~\ref{A:CN} for details of the construction of this equation).  This expression is derived from the concatenated sequence described by Eq.~\ref{Decoupling:Sequence}, and it is simply the product of the largest reduction in strength for each of the $m$ concatenations. This is equivalent to assuming the smallest value of the interaction strength reduction for each of the layers of concatenation, thus providing a lower bound on the final interaction strength. We note that we can focus only on the last half of the chain without loss of generality because the filter strength typically decreases as a function of $d$. 

In Fig.~\ref{Fig:F3}d we numerically evaluate this function for up to $N=2^{20}$, revealing exotic fractal-like features due to a recursive structure in the definition in terms of the binary digits of $Q$.  More importantly, we find that the special structure of our pulse sequences gives a much stronger runtime upper bound with \emph{polynomial scaling}, than that suggested by the naive expectation.  Examining the results, we conjecture that a lower bound for this function is given by $Q^{-2.585}$ (red line, Fig.~\ref{Fig:F3}d), the exponent given by $-\log_2 6$.   This is an exceptional observation, highlighting the practical relevance of this approach.

We may also consider the relevance of control errors in instances where complex concatenated sequences are employed.  A chief source among these is the role of non-instantaneous control may be simply bounded by minimizing the ratio of $\tau_{\pi}/T$ for a given sequence. Since we have demonstrated that the worst-case pulse number required to enact an arbitrary transformation scales polynomially with qubit number, we may therefore also bound the growth of this error.  Error-resistant compensating pulses may also be employed, as has been discussed in the literature~\cite{Khodjasteh:2009, Green:2012}, or recent results assuming bounded-strength controls may be considered~\cite{Viola:2013}.

\section{Efficient Linear Program for Finding Optimal Pulse Sequences}\label{S:lp}


The existence of the universal filtering space guarantees that \emph{some} pulse sequence exists in order to provide a desired transformation.   The process of mapping a desired Hamiltonian for simulation to an appropriate and efficient ``program'' of filters is accomplished as a problem in \emph{linear programming}. The algorithm presented here actually \emph{finds} a valid pulse sequence (if one exists given the input filters) and minimizes the total amount of time spent applying pulses.  Our results show that the problem of compiling pulse sequences can be solved efficiently in $N$, the number of spins, subject to some mild caveats. Additionally, many fast, practical implementations of linear programming algorithms exist to solve this problem at scales of interest for the implementation of useful quantum simulations in polynomial time. 

Consider a set of elementary filter functions $f_1, \ldots,f_m$, which we think of as vectors of dimension $N-1$. These could come from e.g.\ the basic $\Lambda$ and $\Gamma$ filters defined above, together with $k$ levels of concatenation for some fixed $k$. We can also add any other filters so long as we restrict to a polynomial upper bound on $m$, i.e.\ $m= O(N^c)$ for some constant $c$ independent of $N$, in order to ensure that the algorithm below runs in polynomial time. In particular, we can add the $O(N)$ specific ``basis'' we constructed in the Appendix at $k = O(\log N)$ levels of concatenation which we use there to prove the universality of or scheme.

Given this set of dynamic filters, we wish to compile a positive linear combination of pulse sequences that will generate a given Hamiltonian. As mentioned above, this problem can be cast as a \emph{linear program}, which is an efficient method for finding optima of linear objective functions subject to linear equality and inequality constraints~\cite{Boyd2004}. Using standard linear algebra routines, this  procedure may be implemented numerically and provides provably optimal solutions in terms of the total evolution time; the runtime of a linear program with $n$ input variables and $\textrm{poly}(n)$ constraints is polynomial in $n$. 

To see that our problem is a linear program, we make the following observations. Given a desired vector of couplings $\boldsymbol{\Omega}_{\textrm{eff}}$, we wish to find time steps $t_j \ge 0$ such that 
\begin{equation}\label{E:equalityconstraint}
	\sum_{j=1}^m t_j \boldsymbol{f}_j = \frac{\boldsymbol{\Omega}_{\textrm{eff}}}{\Omega} \,.
\end{equation}
Here we actually have a \emph{vector} equation, and the division on the righthand side is done elementwise. Suppose that each filter $\boldsymbol{f}_j$ has a cost $c_j$ associated to it (for simplicity we can imagine that all of these costs are equal.) Then a linear program which will compile a given pulse sequence to generate the effective coupling $\boldsymbol{\Omega}_{\textrm{eff}}$ is 
\begin{equation}
	\min_t\ c^Tt  \quad \text{subject to} \quad \textstyle{\sum_{j=1}^m t_j \boldsymbol{f}_j = \tfrac{\boldsymbol{\Omega}_{\textrm{eff}}}{\Omega}} \ , \ t_j \ge 0\,.
\end{equation}

The form of the program presented above may be converted to the ``standard'' form for linear programming by adding slack variables with additional equality constraints for each of the components of the vector equation.  Finding a solution requires that the selected filter set is chosen to be universal.   Regardless, the program will terminate in time polynomial in $m$ and $N$ and will minimize the total amount of evolution time required to implement the desired coupling.  

Further improvements might be desired, such as having only a few nonzero values for $t_j$ (that is, requiring implementation of only a few different filters). By Carath\'{e}odory's theorem~\cite{Rockafellar1972}, only $N$ of the filters need to be nonzero for any fixed value of $\boldsymbol{\Omega}_{\textrm{eff}}/\Omega$ in order to guarantee a solution. However, finding such sparse solutions is in general NP-hard, so one would likely have to resort to heuristic methods to make improvements along these lines. 

\section{Example programs}\label{S:examples}
Some examples of Hamiltonian engineering will help to reveal the power, generality, and utility of our approach. A simple example is the application of the filter $\bigl[f^{\Lambda}_{k}(d)+I\bigr]$, where $I$ indicates free evolution of equal duration. This particular filter ensures that all qubits separated by distance $k$ will be fully decoupled, combining a period of pulsed modulation with free-evolution, and exploiting the fact that $f^{\Lambda}_{k}(k)=-1$. Using concatenation, we can eliminate all interactions in the spin chain except for, e.g., nearest-neighbor interactions. This Hamiltonian is vital not only for simulation of quantum magnetism, but also many other quantum information protocols \cite{Ronke:2011,Keating:2012}. 

A particularly useful example of a relevant Hamiltonian mapping relates to problems in quantum magnetism~\cite{Fisher72, Sak73, Blote2002} where long-range qubit interactions can be engineered to scale as $\Omega_d\propto d^{-\alpha}$, $\alpha\in[0,3]$: a form of interaction that arises in phonon-mediated spin simulators using trapped ions \cite{Porras:2004,Kim:2009, Britton:2012}. In practice, many simulators cannot reach the achievable limits of this range, or there may be a desire to induce a scaling outside of the range of this native interaction. 


Here we provide details of how, using the pulse sequences we've introduced, one may adjust power-law interactions beyond the accessible bounds native to the underlying hardware.  As an example, consider a system whose qubits interact through a Coulomb interaction such that $\Omega_d\propto\Omega_1/d$.  In order to use this system to simulate a many-body system whose particles interact through a $1/d^2$ potential, we must construct a filter with functional dependence $f(d)\propto1/d$ so that $\Omega_d\to\Omega_d/d$.  Such a filter may obviously be applied repeatedly to modify arbitrary power-law interactions.  

For a system of $N$ qubits, this transformation may be achieved through the sequential application of increasingly complex concatenations of $\Lambda_{N+1}$.  Any $\Lambda_k$, $k>N$ can be used, but $k=N+1$ minimizes the number of pulses in the primitive filter.  In a system of $N$ qubits, the map generated by $\Lambda_{k}$ is linear over the entire chain when $k>N$ so that it can be described as $f^{\Lambda}_{k}(d)=1-2\frac{d}{k}$.  Rewriting this expression as $-\frac{2}{k}\left(d-\frac{k}{2}\right)$ highlights that this is equal to the linear term in the Taylor expansion of $\frac{k/2}{d}$ around $d=k/2$, with a radius of convergence of $k/2$;
\begin{equation}
	\frac{k/2}{d}=\sum_{j=0}^{\infty}\left(\frac{-(d-k/2)}{k/2}\right)^j \,.
\end{equation}

We define the evolution operator $U_{1/d^n}$ which is generated by an interaction that obeys a power law of $\Omega_{d}=\Omega_{1}/d^n$.  Consider the following filtered evolution operator $V$,
\begin{align}
V&=U_{1/d^n}(t)\Lambda_{N+1}(U_{1/d^n}(t))\Lambda_{N+1}^2(U_{1/d^n}(t))\Lambda_{N+1}^3(U_{1/d^n}(t)) \ldots \\
&=\mathrm{exp}\left[-i\sum_{d=1}^N\frac{\Omega_{1}}{d^n}\left(1+f^\Lambda_{N+1}(d)+\left(f^\Lambda_{N+1}(d)\right)^2+...\right)t\sum_{j=1}^{N-d}X_jX_{j+d}\right]\\
&=\mathrm{exp}\left[-i\sum_{d=1}^N\frac{\Omega_{1}\left(N+1\right)}{2d^{n+1}}t\sum_{j=1}^{N-d}X_jX_{j+d}\right]=U_{1/d^{n+1}}.
\end{align}
Here the exponent on the symbol $\Lambda^{x}_{k}$ denotes concatenation to $x$ levels, and the last line follows by summing the geometric series. The effective interaction strength now obeys a new inverse power law where the exponent has been incremented by a single power.   More generally, one may choose different timings for the various filtering operations in order to construct a more general dynamical mapping function,
\begin{align}
V&=U_{1/d^n}(\alpha_0t)\Lambda_{N+1}(U_{1/d^n}(\alpha_1t))\Lambda_{N+1}^2(U_{1/d^n}(\alpha_2t))\Lambda_{N+1}^3(U_{1/d^n}(\alpha_3t))\ldots\\
&=\mathrm{exp}\left[-i\sum_{d=1}^N\frac{\Omega_{1}}{d^n}g(d,\left\{\alpha_i\right\})t\sum_{j=1}^{N-d}X_jX_{j+d}\right]
\end{align}
The function $g(d,\left\{\alpha_i\right\})$ is restricted by the fact that the coefficients $\left\{\alpha_i\right\}$ are necessarily positive.  This implies that the function $f$ must have Taylor expansion coefficients that alternate in sign, meaning that the technique is capable of producing any inverse power law filter but is incapable of producing polynomials with positive exponents.  Another possibility to realize these power laws is to use the linear program of the previous section together with a family of pulse sequences to try to find the most time-efficient power law across all $N$ of the spins. 

\section{Application to Adiabatic Protocols}\label{S:Adiabatic}
Next, we describe how this approach may be applied to adiabatic simulators and preparation of entangled ground-states of designer Hamiltonians~\cite{BabARX}. Say we wish to adiabatically evolve to the ground state of a target Hamiltonian, $H_t$, but can only turn on the available Hamiltonian $H_a$ in a particular experimental apparatus. Returning again to our 1D qubit chain, we initialize in $\ket{g_s}$, which is the ground state of a simple Hamiltonian such as $H_s=-\frac{\omega}{2}\sum_j Z_j$, and allow the system to evolve under the time-dependent Hamiltonian, $H(t)=\left(1-\frac{t}{\tau}\right)H_s+\frac{t}{\tau}H_a$.  If the evolution is adiabatic, then $|\bra{g_a}\hat{U}_{a}(\tau)\ket{g_s}|^2\approx1$ at the end of the interaction.  

Now we apply stroboscopic filtering to drive the system to the new ground state of $H_t$, $\ket{g_t}$. The filtering must be done repeatedly throughout the adiabatic evolution since the Hamiltonian generally does not commute with itself at different times. The rate of application is set such that, to first order, the Hamiltonian is constant over the pulse period, $\Delta t$, giving a total evolution defined by $\prod_{l=1}^R\mathcal{F}\bigl[\hat{U}_{a}(l\Delta t,(l-1)\Delta t)\bigr]$ with $R=\tau/\Delta t$ the number of filtering operations performed during the adiabatic ramp, $\mathcal{F}$ the dynamic filter, and $\hat{U}_a(t_1,t_2)$ the time-ordered evolution from $t_1$ to $t_2$.

\begin{figure}[tb]
\centering
\includegraphics[width=11cm]{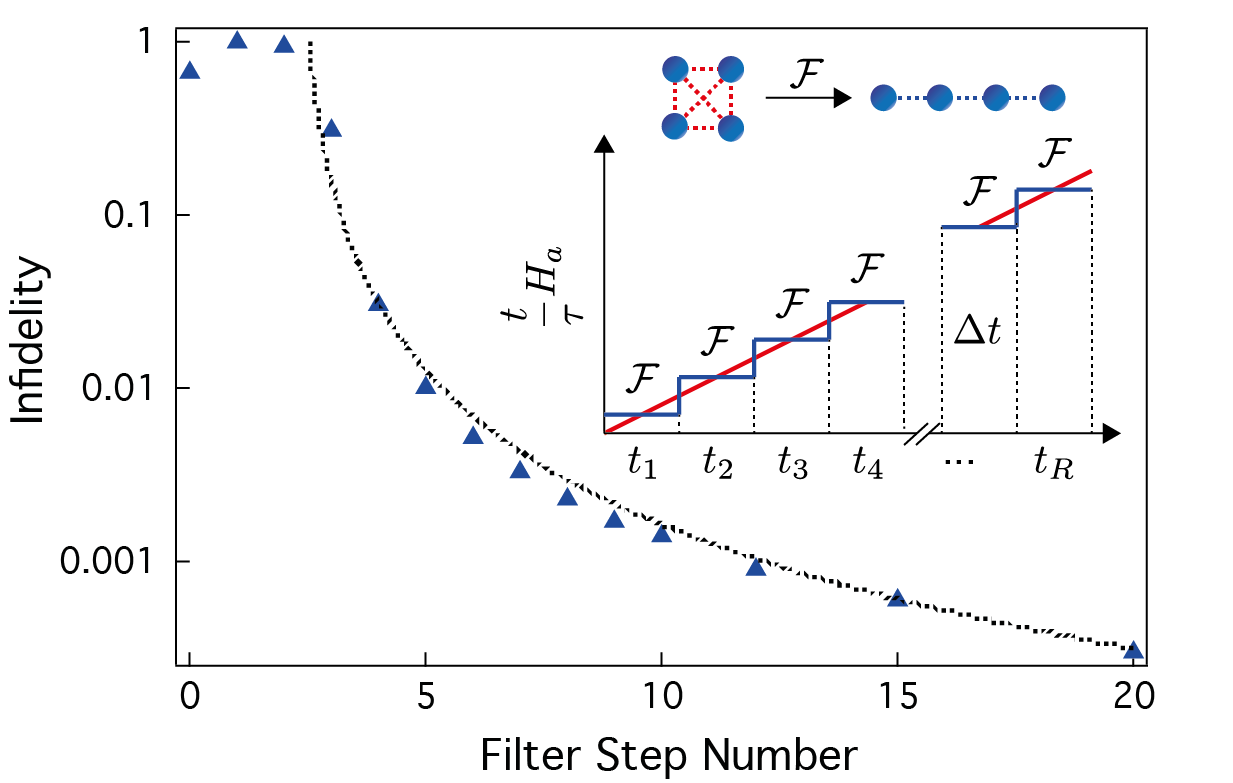}\\
\caption{(color online)  Application of dynamic filtering to the adiabatic evolution of a quantum simulator.  Inset: Schematic of the approximation of breaking a linear ramp of Hamiltonian $H_{a}$ into piecewise-constant segments during which a dynamic filter, $\mathcal{F}$ is applied.  In the example treated here, we filter an all-to-all interaction between four spins to give only nearest-neighbor interactions using $\mathcal{F}(U_0)=\left(\Lambda_2\circ\Lambda_3\right)(U_0)$.  In the main panel we calculate the state fidelity of the adiabatic evolution as a function of the number of filtering operations.  The error,  $\approx\mathrm{Var}(H_f^{(N)})\Delta t^2/4\hbar^2+\mathcal{O}(\Delta t^3)$, where the variance is calculated with the respect to the initialized ground state of $H_s$, decreases quadratically with filtering step.  The dashed line is a guide to the eye showing the quadratic improvement.}
\end{figure}

Using the filtering approach in adiabatic simulators to reach a target ground state $\ket{g_{t}}$, requires repeatedly applying the sequence at a rate fast enough that the evolving Hamiltonian is approximately constant over the time it takes to apply the pulse sequence. Defining a filter $\mathcal{F}$ that maps the evolution operator generated by $H_a$ to one generated by $H_t$ gives a piecewise-constant filtered adiabatic evolution operator 
\begin{equation}
	U_a=\prod_{j=1}^R\mathcal{F}\bigl(\mathrm{exp}\left[-iH(t_j)\Delta t\right]\bigr)\,,
\end{equation}
where $R=\tau/\Delta t$ is the number of filtering operations performed during the adiabatic ramp.  The state then evolves during these discrete timesteps as 
\begin{equation}
	\ket{\psi}=e^{-iH_f^{(R)}\Delta t}e^{-iH_f^{(R-1)}\Delta t}\cdots e^{-iH_f^{(1)}\Delta t}\ket{g_s}\,.
\label{eq:Psi}
\end{equation}
Here $H_f^{(j)}$ is the filtered Hamiltonian at $t=j\Delta t$.  

The error accrued due to the Hamiltonian changing during the filtering operations can be calculated as the overlap of $\ket{\psi}$ with a state $\ket{\phi}$ whose evolution includes the first-order time-dependence of the Hamiltonian during filtering operations.  In the adiabatic approximation, the time evolution operator is given by $\mathrm{exp}[-i\int_{t}^{t+\Delta t} H(t')dt']$ which can be approximated by $\mathrm{exp}[-i( H(t)\Delta t+\frac{1}{2}\dot{H}(t)\Delta t^2)]$ when $\Delta t$ is small.  In cases like ours where, $\hat{P}_j\frac{d}{dt}H(t)\hat{P}_j=\frac{d}{dt}\hat{P}_jH(t)\hat{P}_j$ we can write 
\begin{align}
	\ket{\phi} = \ e^{-i\left(H_f^{(N)}\Delta t+\frac{1}{2}\dot{H}_f^{(N)}\Delta t^2\right)}
		e^{-i\left(H_f^{(N-1)}\Delta t+\frac{1}{2}\dot{H}_f^{(N-1)}\Delta t^2\right)}
		 \cdots 
		 e^{-i\left(H_f^{(1)}\Delta t+\frac{1}{2}\dot{H}_f^{(1)}\Delta t^2\right)}\ket{g_s},
\label{eq:Phi}
\end{align}
which is meant to capture the detrimental effect on the filtering operation due to the changing Hamiltonian to lowest order in $\Delta t$.  The fidelity of the filtered adiabatic protocol can be defined as the overlap of $\ket{\psi}$ and $\ket{\phi}$ as defined in Eqs.~(\ref{eq:Psi}) and (\ref{eq:Phi}). Thus, we have
\begin{align}
\left|\left\langle\phi|\psi\right\rangle\right|^2&=\left|\bra{g_s}e^{i(H^{(1)}\Delta t+\frac{1}{2}\dot{H}^{(1)}\Delta t^2)}\cdots e^{i(H^{(R)}\Delta t+\frac{1}{2}\dot{H}^{(R)}\Delta t^2)}e^{-iH^{(R)}\Delta t}\cdots e^{-iH^{(1)}\Delta t}\ket{g_s}\right|^2\\
&=\left|\bra{g_s}e^{i\frac{1}{2}\sum_{j=1}^R\dot{H}^{(j)}\Delta t^2+\mathcal{O}(\Delta t^3)}\ket{g_s}\right|^2,
\end{align}
by the Baker-Campbell-Hausdorff formula.  The sum $\sum_{j=1}^R\dot{H}^{(j)}\Delta t$ can be approximated by an integral when $\Delta t$ is small enough so that the overlap is,
\begin{align}
\left|\left\langle \phi|\psi\right\rangle\right|^2&\approx \left|\bra{g_s}e^{i\frac{1}{2}\int_{t_1}^{t_f}\dot{H}(t)dt\Delta t}\ket{g_s}\right|^2\\
&=\left|\bra{g_s}e^{i\frac{1}{2}\left(H(t_f)-H(t_1)\right)\Delta t}\ket{g_s}\right|^2.
\end{align}
From this last expression, we see that the error can be thought of as the probability of the Hamiltonian $\frac{1}{2}\left(H(t_f)-H(t_1)\right)$ evolving the system out of the state $\ket{g_s}$ in a time $\Delta t$.  To lowest order in $\Delta t$, this error is $\mathrm{Var}(H(t_f)-H(t_1))\Delta t^2/4\hbar^2$ where the variance is calculated with respect to $\ket{g_s}$.  In the adiabatic evolution described in this article, $\ket{g_s}$ is the ground state of $H(t_1)$, meaning that the expression for the error can be further simplified to $\mathrm{Var}(H(t_f))\Delta t^2/4\hbar^2$.

To test this, we numerically integrate the Schr\"{o}dinger equation for a system of $4$ qubits undergoing adiabatic evolution, starting from a distance-independent spin coupling, $\Omega_{d}\equiv \Omega$, and filtered to produce only nearest-neighbor coupling. We calculate the error accrued due to deviation from the assumption of a piecewise-constant Hamiltonian during the filtering operations as the overlap of the ideal state with a state whose evolution includes a first-order time-dependence of the Hamiltonian during filtering and find that the infidelity decreases approximately as $\left(\Delta t\right)^2$. In this case, the overlap between the ground state of the target Hamiltonian and that of the unfiltered Hamiltonian is $|\langle g_{t}|g_{a}\rangle|^2=0.33$, but with the application of dynamic filters, surpasses $10^{-2}$ after just six filtering steps. 

\section{Moving Beyond Ising Interactions}\label{S:Heisenberg}
We now relax the original constraints we placed on our interaction Hamiltonian in order to simplify the analytic treatments presented above in demonstrating the basic functionality of our approach.  All of these results continue to hold when our Ising-type model is generalised to include both the homogeneous on-site terms and the presence of long-range Heisenberg-type interactions.  

\subsection{Heisenberg Interactions}
Similar to the work of Hodges et al. \cite{Hodges:2007}, an Ising interaction (e.g. $XX$) can be modified into a Heisenberg-type ($S\cdot S$) by including global basis-changing $\pi/2$ pulses in the sequence of applied filters.  Similarly, any native Heisenberg interaction may be modified into any other using pulse sequences applied in the appropriate basis and concatenated

We may rigorously bound the error and runtime as a polynomial function of $T$ and $N$, finding only a constant multiplicative penalty in run-time and Trotter error (see below in \Sref{S:transversefield}). Suppose that our target Hamiltonian is of the form (cf.\ Eq.~\ref{eq:Hamiltonian_aniso})
\begin{equation}
	H = \sum_d \biggl( \Omega^x_d \sum_j X_j X_{j+d} + \Omega^y_d \sum_j Y_j Y_{j+d} + \Omega^z_d \sum_j Z_j Z_{j+d} \biggr)\,.
\end{equation}
For example, if $\Omega_j^v=0$ for $v=x,y,z$ and $j\ge2$, then this is the standard spin-$1/2$ XYZ model. This Hamiltonian can be engineered through a simple extension of the method we describe above.  Assuming the native interaction Hamiltonian consist solely of $X_j$ operators, we build up the time evolution operator using another layer of Trotterization.  To be more precise, we generate the $Y_j$ interactions by applying global $\pi/2$ pulses about the $Z$ axis, and the $Z_j$ interactions are generated by global $\pi/2$ pulses about the $Y$ axis.  If the filtering sequence is represented by the superoperator $\mathcal{F}$, and the global $\pi/2$ pulses are represented by $\mathcal{G}$, the Heisenberg evolution $U_H$ is given by,
\begin{equation}
U_H=\mathcal{F}(U)\mathcal{G}_z\mathcal{F}(U)\mathcal{G}_z\mathcal{G}_y\mathcal{F}(U)\mathcal{G}_y.
\end{equation}
While this method necessarily incurs a Trotter-type error in the implementation, we demonstrate below that this error can be bounded, and provide such a bound for the simulation error (that is, for the case of perfect gates).

\subsection{Inclusion of a Transverse Field\label{S:transversefield}}

Previously, in the case of the Ising coupling, the transverse field was set to zero so that all of the terms in the Hamiltonian would commute with one another. This resulted in an exact mapping from an initial Hamiltonian of the form 
\begin{equation}
	H_I=\sum_d\Omega_d\sum_j X_j X_{j+d}
\end{equation}
to another Hamiltonian of the same form, but with the distance function being modified as $\Omega_d \to \Omega_d f(d)$. The functional dependence $f(d)$ can take an arbitrary form, up to an overall rescaling factor.

To incorporate a homogeneous transverse-field term of the form
\begin{equation}
	H_T=B \sum_j Z_j\,,
\end{equation}
we use a Suzuki-Trotter-type decomposition of the evolution operator~\cite{Trotter1959, Suzuki:1990, Suzuki1991}. We assume that the overall strength $B$ of the transverse field is a tunable parameter so that we can match its rescaled strength relative to the rescaled coupling strength $\tilde{\Omega}_d$ by the same overall scale factor. Given this assumption, there are two natural ways to analyze this situation: an always-on field and a field which can be switched on and off rapidly. 

In some implementations, it is further reasonable to expect that the \emph{interactions} can be switched on and off rapidly (e.g., some trapped ion experiments~\cite{Islam2011}). We will also analyze this case, since in fact it is the easiest to analyze and gives insight into the other results.

\subsection{Bounding the Trotter error in a simulation}

We wish to bound the deviation between the exact and the approximate (Suzuki-Trotter) evolution, assuming that all of our pulse sequences are accurate and perfectly timed. We denote the ideal evolution operator by
\begin{equation}
	V(t)=\mathrm{exp}\left[-i\sum_{j}\hat{P}_j(H_{T}+H_I)\hat{P}_j\Delta t\right]=\mathrm{exp}\left[-i(H_{T}+\tilde{H}_I)t\right],
\end{equation}
where $\left(\sum_{j}\hat{P}_jH_I\hat{P}_j\right)\Delta t=\tilde{H}_It$. The approximate evolution is defined in terms of the second-order integrator (the ``split-step'' method), given by the following formula:
\begin{equation}
	U_r(t) = \left( \prod_{j=1}^m \exp[-i H_j t / 2r] \prod_{j'=m}^1 \exp[-i H_{j'} t / 2r] \right)^r \,.
\end{equation}
Here the $H_j$ are the allowed evolutions over a given time interval, and their exact definition depends on which resources we allow ourselves in the protocol, but they are related to the ideal effective Hamiltonian (here called simply $H$) by
\begin{equation}
	H = H_T + \tilde{H}_I = \sum_{j=1}^m H_j \,.
\end{equation}
We will discuss the specific choices below. We have also introduced a parameter $r$, the Trotter number, which counts the number of steps in our Suzuki-Trotter expansion and hence controls our error. 

Note that we \emph{cannot} use the higher-order integrators which provide faster convergence because they all require evolution for negative values of $t$~\cite{Suzuki1991}. If rapidly flipping the sign of the interactions and the transverse field is feasible in a given system, then these higher-order integrators could be used to give sharper error bounds than the ones given below. 

Given this very general framework, we can upper bound the error as follows~\cite{Berry:2007}
\begin{equation}\label{E:errorbound}
	\| V(t) - U_r(t) \| \le 16 m^3 \|H\|^3 t^3 / r^2 \,.
\end{equation}
Here the error is in terms of the operator norm (largest singular value). Note that $\|H\| \le N^2/2 + B N$ by the triangle inequality, so this quantity is always polynomial in the number of spins. This bound on the norm of the difference has an operational meaning: it gives an upper bound on the trace distance between the evolved states, namely
\begin{equation}
	\|V-U\| \ge \frac{1}{2} \mathrm{Tr}\bigl| V\rho V^\dag - U\rho U^\dag\bigr|
\end{equation}
for any initial state $\rho$ and unitaries $U$ and $V$. 

To make the error bound in Eq.~(\ref{E:errorbound}) more concrete, we need to determine $m$ for a given protocol. The form of the bound above indicates that a smaller number of different Hamiltonians in use results in improved error scaling.  

The simplest case to consider is that where both the interactions and the transverse field can be switched on and off rapidly. In this case we can use the above bound with $m = 2$, in which we simply switch between (1) the exact evolution of the effective Hamiltonian with zero transverse field, and (2) the transverse field alone with no interactions. That is, we have 
\begin{equation}
	H_1 = \tilde{H}_I \quad \text{and} \quad H_2 = H_T \,.
\end{equation}
Note that $\tilde{H}_I$ can be implemented exactly by subdividing each Suzuki-Trotter step into smaller segments, as discussed in the main text. 

Next, consider the case where the interactions are always on, but the transverse field can be switched on and off rapidly. Here a good strategy is to turn on the transverse field in short, strong bursts, and turn it off while applying the filter functions. The control Hamiltonians during each of the $m=2$ Suzuki-Trotter steps are given by
\begin{equation}
	H_1 =  \tilde{H}_I \quad \text{and} \quad H_2 = \delta (H_I + \tfrac{1}{\delta} H_T) \,.
\end{equation}
Here we introduce an additional parameter $\delta \ll 1$, and we incur a small error in the final Hamiltonian,
\begin{equation}
	H_\delta =  \tilde{H}_I + H_T + \delta H_I \,.
\end{equation}
We can choose $\delta$ as small as possible to be compatible with the assumption of accurate and rapid switching of the transverse field. Thus, the error bound is nearly the same as the previous $m=2$ bound, except that we add a small error from the inexact decomposition.
\begin{align}
	\|V(t) - U_r(t) \| \le \|V(t) - V_\delta(t) \| + \|V_\delta(t) - U_r(t) \| \le 128 \|H_\delta\|^3 t^3 / r^2 + O\bigl(\delta \|H_I\| t\bigr) \,.
\end{align}

Next, consider the most pessimistic case where both the interactions and the transverse field are not rapidly switchable. In this case the best strategy seems to be to use a different Suzuki-Trotter step for each increment in a pulse sequence. In this case, we have no choice but to use the bound in Eq.~(\ref{E:errorbound}) by setting $m$ equal to the total number of pulses in the pulse sequence. For a chain of length $N$, we can always bound the maximum number of pulses required by $m \le N$, which follows from Carath\'{e}odory's theorem~\cite{Rockafellar1972}. This gives a pessimistic but nonetheless polynomial bound on the required Trotter number to achieve a fixed error. 

Finally, we remark that the same methods for bounding the Trotter error in this section also apply in \Sref{S:Heisenberg}.  Since this more general case includes, at most, three times the number of pairwise coupling terms in the Hamiltonian, this implies that the error remains polynomial in the number of qubits.

\section{Conclusion}\label{S:conclusion}
In conclusion, we have provided a general framework for dynamic filtering of spin Hamiltonians that enables programmable quantum simulators using single-qubit Pauli operations and a native long-range spin coupling. The scheme we introduce provides a system designer with the ability to perform numerical decomposition of a large class of realizable couplings into the basis of available filters, thus providing a prescription of how to ``program'' the desired interactions. We have further showed how this technique can be applied to augment the adiabatic evolution of a spin Hamiltonian in a form of hybrid adiabatic quantum simulator.   

We are excited by the fact that our approach to combine such filters in realizing a desired program for quantum simulation represents a contribution to a growing body of literature focused on the benefits of functional analysis in the context of protocol, algorithm, and circuit development for quantum information.  In particular, following completion of this work we discovered significant similarity between our approach and the synthesis of diagonal operators for circuit-model quantum computation~\cite{Welch2014}.  Given these connections and the breadth of applicability of functional analysis in quantum information, we hope that our results enable new capabilities well beyond quantum simulation.

The results presented herein demonstrate the utility of the pulse sequences we define for efficiently modifying native Hamiltonians in quantum simulators.  From the perspective of control engineering these results demonstrate that via the quantum control protocols we have defined, the 1D translationally invariant quantum simulator in which we are interested is \emph{efficiently controllable}, requiring resources scaling only polynomially in time and energy (using pulse number as a proxy).  Assuming only Ising-type long range interactions, we may realise a much broader class of Hamiltonians including spin-$1/2$ $XYZ$ Heisenberg-type models with homogeneous on-stie terms and much more exotic multi-axis spin-coupling Hamiltonians with bounded error and polynomial resources.  All of our pulse sequences and methods of filter combination may be applied in order to achieve these unusual models with the small associated penalties in time and resources.

While we have provided explicit examples of how one might tune the power-law of a long-range spin coupling or completely cancel undesired spin couplings on a 1D lattice, it is an interesting open problem to find additional examples where our method can be applied efficiently.  This includes, for instance, the efficient generation of an effective Heisenberg Hamiltonian using only Ising couplings and single-qubit control such as that in Ref.~\cite{Hodges:2007}, or generalizations of our method to systems with a natural 2D geometry of spins. A fundamental open problem is also finding optimal primary filter constructions that minimize the overhead in evolution time for a desired effective Hamiltonian. In addition we wish to explore the possibility of finding a rigorous proof of polynomial runtime scaling, rather than relying only on numeric evidence supporting our claim.  Finally, it is an exciting open question to explore how these techniques may be coupled with notions of fault-tolerance in quantum simulation. We note that after completion of this work we became aware of the use of similar language of Hamiltonian ``filtering'' for quantum simulation, treating quantum simulation and spin transport in a related framework~\cite{Cappellaro:2013}.

\textit{Acknowledgements:}  The authors thank L.~Viola for useful discussions. This work partially supported by the US Army Research Office under Contract Number  W911NF-11-1-0068, the Australian Research Council Centre of Excellence for Engineered Quantum Systems CE110001013, the Office of the Director of National Intelligence (ODNI), Intelligence Advanced Research Projects Activity (IARPA), through the Army Research Office, and the Lockheed Martin Corporation. STF Acknowledges support from an ARC Future Fellowship.  All statements of fact, opinion or conclusions contained herein are those of the authors and should not be construed as representing the official views or policies of IARPA, the ODNI, or the U.S. Government. 

\section*{References}

\bibliography{DynamicalMappingBib}

\bigskip\bigskip
\appendix
\section{Construction of \texorpdfstring{$\C_N$}{C\_N} and proof of universality}\label{A:CN}

Here we provide a proof of the existence of a universal filter space for an $(N-1)$-dimensional interaction space appropriate for $N$ qubits.  In order to construct a suitable set of extreme points to ensure that their convex hull contains the origin and a hypercube around it, we show that it is possible to generate a complete set of effective of interactions such that only qubits that are separated by a specific distance possess nonzero interaction, i.e. $\Omega_{d'}\neq0$ iff $d'=d$, for a single prescribed $d$. These basis interactions, to which we refer as the Kronecker Delta interaction vectors, would be of the form $\alpha (0,0,\ldots,0,\pm1,0,\ldots,0)$ across the $N-1$ dimensions, where $\alpha$ is a positive number, and the $\pm$ sign permits the realisation of either ferromagnetic or antiferromagnetic interactions.  While large $\alpha$ is desirable from the practical perspective of reducing necessary operation times, the actual value does not impact a proof of the existence of a universal filter space.  We will use this basis exclusively in order to present our proof of universal filtering for $N$ qubits, but more efficient approaches to the construction of $\C_{N}$ likely exist.

We proceed by showing how to construct both a positive and negative Kronecker delta interaction vector for the special cases $d=1,2,3,4$ and finally for the generic case $4<d<N$.  The $d=1$ negative Kronecker delta interaction uses $\Lambda_1$ as its base function, which takes value $f_{1}^{\Lambda}(d)=-1$ at $d=1$.  We eliminate all other couplings through a judicious choice of concatenations which are enumerated through the expression, valid for all $1<d<2^m-1$,
\begin{equation}
\left(f^{\Lambda}_2+f^{\Lambda}_0\right)\prod_{n=2}^{m}\left(f^{\Lambda}_{2^{n}}+f^{\Lambda}_{0}\right)\left(f^{\Lambda}_{2^{n}}+\left(1-\frac{1}{2^{n-1}}\right)f^{\Lambda}_{0}\right)=0.
\end{equation}
This statement can be proved by induction with the base case of $m=2$ being easily checked by hand.  

For the case of an arbitrarily long chain of qubits, we now explicitly describe the inductive step.  Assuming that we have set to zero all couplings up to $2^{n+1}-1$ using the concatenation scheme described above, we must show that subsequent concatenation with $(\Lambda_{2^{n+1}}+\Lambda_0)$ and $(\Lambda_{2^{n+1}}+(1-\frac{1}{2^{n}})\Lambda_0)$ will extend the decoupling range out to $d<2^{n+2}-1$.  Using the explicit form for $f^{\Lambda}_k(d)$, it is trivial to show that these additional concatenations zero out the $2^{n+1}-1$, $2^{n+1}$ and $2^{n+1}+1$ couplings.  Additionally, every filter used in the preceeding concatenations takes value $1$ at $d=2^{n+1}$ (since they all have a periodicity equal to a power of $2$), implying that their behavior from $1<d<2^{n+1}$ will be repeated for $2^{n+1}<d<2^{n+2}$.  Given the assumption of the inductive step, this implies that the concatenation has extended the full decoupling to $d<2^{n+2}-1$.  

Constructing the positive Kronecker delta interaction for $d=1$ is accomplished in the same manner, in this case beginning with $\Lambda_0$ as the base function.  In this rubric, a chain of $N$ qubits requires $2m-1=2 \left\lceil \mathrm{log}_2(N+1)\right\rceil-3$ total concatenations for the Kronecker delta interaction at $d=1$.

The negative $d=2$ Kronecker delta interaction begins with $\Lambda_2$ as the base function followed by concatenation with $(\Lambda_1+\Lambda_0)$ to eliminate all interactions for $d$ odd.  The task of eliminating residual interactions maps directly onto the construction of the $d=1$ Kronecker delta interaction, rescaling all distances by a factor of $2$, using the following:
\begin{align}
\left(f^{\Lambda}_1+f^{\Lambda}_0\right)\left(f^{\Lambda}_4+f^{\Lambda}_0\right)\prod_{n=3}^{m}\left(f^{\Lambda}_{2^{n}}+f^{\Lambda}_{0}\right)\left(f^{\Lambda}_{2^{n}}+\left(1-\frac{1}{2^{n-2}}\right)f^{\Lambda}_{0}\right)=0~\forall~d\neq 2.
\end{align}
This scheme requires a total of $2 \left\lceil \mathrm{log}_2(N)\right\rceil-4$ concatenations for the $d=2$ Kronecker delta interaction so that $2m-2=\left\lceil \mathrm{log}_2(N-1)\right\rceil-2$.

The $d=3$ Kronecker delta interaction is constructed starting with $\Lambda_3$ and concatenating with $(\Gamma_3+\Lambda_0/3)$ to decouple all qubits except those that are separated by multiples of three lattice spacings.  The remaining task can be accomplished by a similar sequence of filters indicated by the expression,
\begin{align}
\left(f^{\Gamma}_3+\frac{1}{3}f^{\Lambda}_0\right)\prod_{n=0}^{m}\left(f^{\Lambda}_{9\cdot2^{n}}+f^{\Lambda}_{0}\right)\left(f^{\Lambda}_{9\cdot2^{n}}+\left(1-\frac{1}{3\cdot2^{n-1}}\right)f^{\Lambda}_{0}\right)=0~\forall~d\neq 3,
\end{align}
requiring $2m+3=2\left\lceil \mathrm{log}_2(\frac{N+2}{9})\right\rceil+1$ concatenations.  

The $d=4$ Kronecker delta interaction construction starts with $\Lambda_4$ and proceeds via concatenation with $\Gamma_4$ to decouple qubits that are \emph{not} separated by a multiple of four lattice spacings.  The decoupling is accomplished via the concatenation of the filters indicated by,
\begin{equation}
f^{\Gamma}_4\prod_{n=0}^{m}\left(f^{\Lambda}_{12\cdot2^{n}}+f^{\Lambda}_{0}\right)\left(f^{\Lambda}_{12\cdot2^{n}}+\left(1-\frac{1}{3\cdot2^{n-1}}\right)f^{\Lambda}_{0}\right)=0~\forall~d\neq 4,
\end{equation}
requiring $2m+3=2\left\lceil \mathrm{log}_2(\frac{N+3}{12})\right\rceil+1$ concatenations.

The final construction is for the Kronecker delta interaction at $d=Q$ where $4<Q\leq N-1$, employing $\Gamma_Q$ which is positive for all $d$.  For $d<Q$, $f^{\Gamma}_Q(d)=1-\frac{4}{Q}$.  Applying $((1-\frac{4}{Q})\Lambda_{2^m}+\Gamma_Q)$ for $m=0,1,2,3.....\leq \mathrm{log}_2N$ decouples all qubits except those that are separated by multiples of $Q$ lattice spacings.  The next task of decoupling all qubits separated by multiples of $Q$ can again be mapped onto our original construction of the $d=1$ Kronecker delta interaction by rescaling the distance by $Q$.  The entire decoupling sequence is indicated by the following expression which is valid for all $d\neq Q$, 
\begin{align}
\left(\prod_{n=0}^{m}\left(f^{\Lambda}_{2^{n}}+\frac{1}{(1-\frac{4}{Q})}f^{\Gamma}_{Q}\right)\right)\left(f^{\Lambda}_{2Q}+f^{\Lambda}_0\right)\prod_{q=2}^{p}\left(f^{\Lambda}_{2^qQ}+f^{\Lambda}_0\right)\left(f^{\Lambda}_{2^qQ}+\left(1-\frac{1}{2^{q-1}}\right)f^{\Lambda}_0\right)=0\,
\label{Decoupling:Sequence}
\end{align}
resulting in a total of $m+2p=\left\lceil \mathrm{log}_2\left(N-1\right)\right\rceil+2\left\lceil \mathrm{log}_2\left(\frac{N-1}{Q}\right)\right\rceil$ filtering operations.

We thus see that it is possible to construct the Kronecker delta interaction with $O(\log_{2}(N))$ concatenations.  Accordingly we know that it is possible to define a convex hull that includes the origin and the extremal points defined by the Kronecker deltas above, proving universality for an $N$-qubit chain as defined herein.

The logarithmic scaling we present here is much better than brute force, however the effective coupling strengths that result from $\log N$ levels of concatenation shrink polynomially with $N$, leading to decreased simulation efficiency for large values of $N$ in this particular construction.  An obvious open question is if this universal filtering space can be constructed with only a \textit{constant} number of concatenations, which would be a very strong result.

\end{document}